\documentclass[12pt]{article}

\usepackage{cite}
\usepackage{graphicx}
\usepackage{pict2e}
\usepackage[all]{xy}

\author{Yu.~M.~Zinoviev
       \thanks{E-mail address: Yurii.Zinoviev@ihep.ru} \\[0.5cm]
        {\it Institute for High Energy Physics} \\
        {\it of National Research Center "Kurchatov Institute"} \\
        {\it Protvino, Moscow Region, 142280, Russia}}
\title{On the frame-like multispinor formalism \\
for massive higher spins in $d=4$}

\date{}

\begin{document}

\maketitle

\begin{abstract}
In this paper, we fill some gap in the existing literature on higher
spins by presenting an explicit solution to the on-shell constraints
for a frame-like, gauge invariant description of massive, higher spin
fields in $d=4$. We begin with the massive spin 2 and massive spin 5/2
as simple illustrations, and then consider arbitrary integer
and half-integer spin. We also show that our results allow
us to find explicit solutions to the so-called unfolded equations
that determine all higher-order derivatives of the physical field
that are non-zero on-shell.
\end{abstract}

\thispagestyle{empty}
\newpage
\setcounter{page}{1}

\section{Introduction}

The frame-like gauge invariant formalism is a powerful tool for
describing  massive higher spin fields and their possible
interactions \cite{Zin08b,PV10,KhZ19,Zin24a}. As the very important
ingredients it contains so-called extra fields that do not appear in
the free Lagrangian but allow us to describe higher derivative
interactions typical for higher spins in a compact and elegant way
(see e.g. resent results in \cite{Zin26}). To establish a connection
of the extra fields with the Lagrangean fields we need the appropriate
on-shell constraints\footnote{In the metric-like formalism
one considers the terms  on-shell and on equation of motion as the
synonyms. But it is not so in the frame-like formalism.}. From one
hand, these constraints must contain Lagrangean equations for the
physical fields; from the other hand, they must allow us to express
all extra fields in terms of derivatives of physical fields. Such
constraints exist and lead to the correct number of physical degrees
of freedom. However, as far as we know, the explicit solutions for the
extra fields have not been presented till now\footnote{The situation
here is similar to the one in frame formalism for gravity. There
exists a well-known torsion zero condition that allows one to express
Lorentz connection as a derivative of the frame field. But in many
calculations it appears to be sufficient directly use that torsion is
zero.}. Our goal in this note is to provide such solutions. To
simplify presentation we restrict ourselves to the flat four
dimensional case.

One of the nice features of the frame-like gauge invariant formalism
is that each field has a corresponding gauge invariant object 
(two-form for the gauge field and one-form for a Stueckelberg one)
which we call curvature. In-particular, they make it possible to
formulate on-shell constraints independently of any gauge choice.
Recall that in the gauge invariant formalism there is a one-to-one
correspondence between gauge and Stueckelberg fields, as it is quite
natural, because in the massive case all gauge symmetries are
spontaneously broken. This allows us to use the so-called unitary
gauge, where all Stueckelberg fields are set to zero. Such a choice
greatly simplifies the solution of on-shell constraints; and as a
by-product we confirm that they lead to the correct number of physical
degrees of freedom.

However, to describe all possible interactions we need even higher
derivatives (see e.g. \cite{TV24}). A very effective way to describe
all higher derivatives of the physical fields that do not vanish
on-shell is using the so-called unfolded equations\footnote{Here we
restrict ourselves to the massive case; for the massless one see
e.g.\cite{BUV21} and references therein}
\cite{PV10,BPSS15,Zin15,BSZ16,KhZ19,KhZ20,Zin24}. Using our results
for the extra fields as input, we find solutions for the corresponding
unfolded equations as well.

Our paper is organized as follows. In Sections 2 and 3 we present our
results for massive spin 2 and massive spin 5/2, as the simplest
bosonic and fermionic examples. Sections 4 and 5 deal with
arbitrary integer and half integer spins. An appendix contains
some general remarks on the unfolded equations.

\noindent
{\bf Notation and conventions} We work in the multispinor version
of the frame-like gauge invariant formalism \cite{KhZ19} where all
objects are forms with some number of completely symmetric dotted and
un-dotted spinor indices $\Phi^{\alpha(k)\dot\alpha(l)}$, 
$\alpha,\dot\alpha = 1,2$. For the coordinate free description of the
flat four dimensional space we use a background  frame
$e^{\alpha\dot\alpha}$ and background Lorentz covariant derivative $D$
such that
\begin{equation}
D \wedge e^{\alpha\dot\alpha} = 0, \qquad
D \wedge D = 0.
\end{equation}
We also use the basic two-forms $E^{\alpha(2)}$ and
$E^{\dot\alpha(2)}$ defined as follows
\begin{equation}
e^{\alpha\dot\alpha} \wedge e^{\beta\dot\beta} =
\epsilon^{\alpha\beta} E^{\dot\alpha\dot\beta} +
\epsilon^{\dot\alpha\dot\beta} E^{\alpha\beta}.
\end{equation}
Similarly to the three dimensional case \cite{KP18,KhZ22}, we use the
possibility to express any one-form as a combination of zero-forms
\begin{equation}
\Phi^{\alpha(k)\dot\alpha(l)} = e_{\beta\dot\beta}
\phi^{\alpha(k)\beta\dot\alpha(l)\dot\beta} + e_\beta{}^{\dot\alpha}
\phi^{\alpha(k)\beta\dot\alpha(l-1)} + e^\alpha{}_{\dot\beta}
\phi^{\alpha(k-1)\dot\alpha(l)\dot\beta} + e^{\alpha\dot\alpha}
\phi^{\alpha(k-1)\dot\alpha(l-1)},
\end{equation}
which corresponds to decomposition into irreducible Lorentz group
representations. As we will see in what follows, on-shell we obtain
\begin{equation}
\Phi^{\alpha(k)\dot\alpha(l)} \approx e_{\beta\dot\beta}
\phi^{\alpha(k)\beta\dot\alpha(l)\dot\beta}
\end{equation}
and this greatly simplifies calculations.

\section{Massive spin 2}

A massive spin 2 in four dimensions has five helicities 
$(\pm 2, \pm 1, 0)$ so its frame-like gauge invariant description  
requires three physical fields: one-forms $H^{\alpha\dot\alpha}$, $A$
and a zero-form $\varphi$ as well as three auxiliary fields (that are
necessary for the Lagrangian being of first order in derivatives):
one-form $\Omega^{\alpha(2)} + h.c.$ and zero-forms 
$B^{\alpha(2)} + h.c.$,  $\pi^{\alpha\dot\alpha}$. Note that there is
a one-to-one correspondence between the gauge fields (one-forms) and
the Stueckelberg fields (zero-forms), as is natural in a massive case
were all gauge symmetries are spontaneously broken and so each gauge
field has its own Stueckelberg partner. The Lagrangian has the
following form\footnote{This structure is typical for the gauge
invariant Lagrangians describing massive fields: the sum of kinetic
and mass-like terms for all helicities plus cross terms (third line)
gluing them together.}:
\begin{eqnarray}
{\cal L}_0 &=& \Omega^{\alpha\beta} E_\beta{}^\gamma
\Omega_{\alpha\gamma} + 2 \Omega^{\alpha\beta} e_\beta{}^{\dot\alpha}
D H_{\alpha\dot\alpha} + 4E B_{\alpha(2)} B^{\alpha(2)} + 2
E_{\alpha(2)} B^{\alpha(2)} D A + h.c. \nonumber \\
 && - 6E \pi_{\alpha\dot\alpha} \pi^{\alpha\dot\alpha} - 12
E_{\alpha\dot\alpha} D \varphi \nonumber \\
 && + 2M [ \Omega^{\alpha(2)} E_{\alpha(2)} A - 2 B^{\alpha\beta}
E_\beta{}^{\dot\alpha} H_{\alpha\dot\alpha} + h.c.] + 12M 
E_{\alpha\dot\alpha} \pi^{\alpha\dot\alpha} A \nonumber \\
 && + 2M^2 [ H^{\alpha\dot\alpha} E_\alpha{}^\beta H_{\beta\dot\alpha}
+ h.c.] + 12M^2 E_{\alpha\dot\alpha} H^{\alpha\dot\alpha} \varphi +
12M^2 E \varphi^2
\end{eqnarray}
This Lagrangian is invariant under the following gauge
transformations:
\begin{eqnarray}
\delta \Omega^{\alpha(2)} &=& D \eta^{\alpha(2)} + \frac{M^2}{2}
e^\alpha{}_{\dot\alpha} \xi^{\alpha\dot\alpha} \nonumber \\
\delta H^{\alpha\dot\alpha} &=& D \xi^{\alpha\dot\alpha} + 
e_\beta{}^{\dot\alpha} \eta^{\alpha\beta} + e^\alpha{}_{\dot\beta}
\eta^{\dot\alpha\dot\beta} + M e^{\alpha\dot\alpha} \xi \nonumber \\
\delta B^{\alpha(2)} &=& - M \eta^{\alpha(2)}, \qquad
\delta A = D \xi + M e_{\alpha\dot\alpha} \xi^{\alpha\dot\alpha} \\
\delta \pi^{\alpha\dot\alpha} &=& - M^2 \xi^{\alpha\dot\alpha}, \qquad
\delta \varphi = - M \xi \nonumber
\end{eqnarray}
In the gauge invariant formalism each field has its own gauge
invariant object (curvature). For the gauge fields they are two-forms:
\begin{eqnarray}
{\cal R}^{\alpha(2)} &=& D \Omega^{\alpha(2)} + M E^\alpha{}_\beta
B^{\alpha\beta} + \frac{M^2}{2} e^\alpha{}_{\dot\alpha}
H^{\alpha\dot\alpha} + 2M^2 E^{\alpha(2)} \varphi, \nonumber \\
{\cal T}^{\alpha\dot\alpha} &=& D H^{\alpha\dot\alpha} + 
e_\beta{}^{\dot\alpha} \Omega^{\alpha\beta} + e^\alpha{}_{\dot\beta}
\Omega^{\dot\alpha\dot\beta} + M e^{\alpha\dot\alpha} A, \\
{\cal A} &=& D A + 2(E_{\alpha(2)} B^{\alpha(2)} + E_{\dot\alpha(2)}
B^{\dot\alpha(2)}) + M e_{\alpha\dot\alpha} H^{\alpha\dot\alpha},
\nonumber
\end{eqnarray}
while for the Stueckelberg fields they are one-forms: 
\begin{eqnarray}
{\cal B}^{\alpha(2)} &=& D B^{\alpha(2)} + M \Omega^{\alpha(2)} +
\frac{M}{2} e^\alpha{}_{\dot\alpha} \pi^{\alpha\dot\alpha}, \nonumber
\\
\Pi^{\alpha\dot\alpha}  &=& D \pi^{\alpha\dot\alpha} + M 
(e_\beta{}^{\dot\alpha} B^{\alpha\beta} + e^\alpha{}_{\dot\beta}
B^{\dot\alpha\dot\beta}) + M^2 H^{\alpha\dot\alpha} + M^2
e^{\alpha\dot\alpha} \varphi, \\
\Phi &=& D \varphi + e_{\alpha\dot\alpha} \pi^{\alpha\dot\alpha}
+ M A. \nonumber
\end{eqnarray}
In this formalism the so-called on-shell constraints (analogue of the
torsion zero condition in gravity) play an important role: they
guarantee the correct number of physical degrees of freedom and also
allow one to express all auxiliary fields in terms of the derivatives
of the physical ones. For the case at hands, they look like:
\begin{equation}
{\cal T}^{\alpha\dot\alpha} \approx 0, \qquad
{\cal A} \approx 0, \qquad \Phi \approx 0,
\end{equation}
\begin{eqnarray}
0 &\approx& {\cal R}^{\alpha(2)} - E_{\beta(2)} W^{\alpha(2)\beta(2)},
 \nonumber \\
0 &\approx& {\cal B}^{\alpha(2)} - e_{\beta\dot\alpha}
B^{\alpha(2)\beta\dot\alpha}, \\
0 &\approx& \Pi^{\alpha\dot\alpha} - e_{\beta\dot\beta}
\pi^{\alpha\beta\dot\alpha\dot\beta}, \nonumber
\end{eqnarray}
where $W^{\alpha(4)}$, $B^{\alpha(3)\dot\alpha}$, 
$\pi^{\alpha(2)\dot\alpha(2)}$ are gauge invariant zero-forms which 
are just the first representatives of infinite chains of  gauge
invariant zero-forms satisfying the so-called unfolded equations 
$(0 \le k < \infty)$:
\begin{eqnarray}
0 &=& D W^{\alpha(4+k)\dot\alpha(k)} - e_{\beta\dot\beta}
W^{\alpha(4+k)\beta\dot\alpha(k)\dot\beta} 
+ \frac{4M}{(k+4)(k+5)} e^\alpha{}_{\dot\beta} 
B^{\alpha(3+k)\dot\alpha(k)\dot\beta} \nonumber \\
 && + \frac{M^2}{(k+1)(k+4)} e^{\alpha\dot\alpha} 
W^{\alpha(3+k)\dot\alpha(k-1)}, \nonumber \\
0 &=& D B^{\alpha(3+k)\dot\alpha(k+1)} -  e_{\beta\dot\beta}
B^{\alpha(3+k)\beta\dot\alpha(k+1)\dot\beta}
 + \frac{M}{(k+1)(k+2)} e_\beta{}^{\dot\alpha} 
W^{\alpha(3+k)\beta\dot\alpha(k)} \label{eq_2m} \\
 && + \frac{3M}{(k+3)(k+4)} e^\alpha{}_{\dot\beta} 
\pi^{\alpha(2+k)\dot\alpha(1+k)\dot\beta}
 + \frac{k(k+5)M^2}{(k+1)(k+2)(k+3)(k+4)} 
e^{\alpha\dot\alpha} B^{\alpha(2+k)\dot\alpha(k)}, \nonumber \\
0 &=& D \pi^{\alpha(2+k)\dot\alpha(2+k)} - e_{\beta\dot\beta}
\pi^{\alpha(2+k)\beta\dot\alpha(2+k)\dot\beta}
 + \frac{2M}{(k+2)(k+3)} e_\beta{}^{\dot\alpha} 
B^{\alpha(2+k)\beta\dot\alpha(k+1)} \nonumber 	\\
 && + \frac{2M}{(k+2)(k+3)} e^\alpha{}_{\dot\beta} 
B^{\alpha(1+k)\dot\alpha(2+k)\dot\beta}
 + \frac{k(k+5)M^2}{(k+2)^2(k+3)^2} e^{\alpha\dot\alpha}
\pi^{\alpha(1+k)\dot\alpha(1+k)}. \nonumber
\end{eqnarray}
where coefficient are determined by their self-consistency (see
Appendix).

To simplify an analysis of the on-shell conditions we chose a
so-called unitary gauge where all Stueckelberg fields are set to zero:
\begin{equation}
B^{\alpha(2)} = 0, \qquad 
\pi^{\alpha\dot\alpha} = 0, \qquad
\varphi = 0.
\end{equation}
Then the on-shell conditions on the one-forms give:
\begin{equation}
\Omega^{\alpha(2)} \approx e_{\beta\dot\alpha} 
\omega^{\alpha(2)\beta\dot\alpha} \qquad
H^{\alpha\dot\alpha} \approx e_{\beta\dot\beta}
h^{\alpha\beta\dot\alpha\dot\beta}, \qquad
A \approx 0. \label{sol_1}
\end{equation}
where $\omega^{\alpha(3)\dot\alpha}$ and $h^{\alpha(2)\dot\alpha(2)}$
are zero-forms, while the conditions on the two-forms become
\begin{eqnarray}
{\cal R}^{\alpha(2)} &=& D \Omega^{\alpha(2)} + \frac{M^2}{2}
e^\alpha{}_{\dot\alpha} H^{\alpha\dot\alpha} \approx 
E_{\beta(2} W^{\alpha(2)\beta(2)}, \nonumber \\
{\cal T}^{\alpha\dot\alpha} &=& D H^{\alpha\dot\alpha} + 
e_\beta{}^{\dot\alpha} \Omega^{\alpha\beta} + e^\alpha{}_{\dot\beta}
\Omega^{\dot\alpha\dot\beta} \approx 0, \\
{\cal A} &=& 0. \nonumber 
\end{eqnarray}
Now using (\ref{sol_1}) we obtain
\begin{equation}
{\cal T}^{\alpha\dot\alpha} \approx \frac{1}{2} E_{\beta(2)} 
D^\beta{}_{\dot\beta} h^{\alpha\beta\dot\alpha\dot\beta} -
E_{\beta(2)} \omega^{\alpha\beta(2)\dot\alpha} + h.c.
\end{equation}
A decomposition into irreducible Lorentz group representations (i.e.
symmetrization and anti-symmetrization on spinor indices) gives
\begin{eqnarray}
0 &\approx& \frac{1}{3} D^\alpha{}_{\dot\beta} 
h^{\alpha(2)\dot\alpha\dot\beta} -  \omega^{\alpha(3)\dot\alpha},
\nonumber \\
0 &\approx& D_{\beta\dot\beta} h^{\alpha\beta\dot\alpha\dot\beta}.
\end{eqnarray}
Similarly we obtain
\begin{equation}
{\cal R}^{\alpha(2)} \approx \frac{1}{2} E_{\beta(2)} 
D^\beta{}_{\dot\alpha} \omega^{\alpha\beta(2)\dot\alpha} + \frac{1}{2}
E_{\dot\alpha(2)} D_\beta{}^{\dot\alpha} 
\omega^{\alpha(2)\beta\dot\alpha} - M^2 E_{\dot\alpha(2)} 
h^{\alpha(2)\dot\alpha(2)}
\end{equation}
and this gives
\begin{eqnarray}
0 &\approx& \frac{1}{4} D^\alpha{}_{\dot\alpha} 
\omega^{\alpha(3)\dot\alpha} - W^{\alpha(4)}, \nonumber \\
0 &\approx& D_{\beta\dot\alpha} \omega^{\alpha(2)\beta\dot\alpha}, \\
0 &\approx& \frac{1}{2} D_\beta{}^{\dot\alpha}
\omega^{\alpha(2)\beta\dot\alpha} - M^2 h^{\alpha(2)\dot\alpha(2)}.
\nonumber
\end{eqnarray}
From the last equation it follows that 
$(D^2 = D_{\alpha\dot\alpha} D^{\alpha\dot\alpha})$
\begin{equation}
(\frac{1}{2} D^2 + M^2)h^{\alpha(2)\dot\alpha(2)} \approx 0.
\end{equation}
In combination with the transversality condition
$D_{\beta\dot\beta} h^{\alpha\beta\dot\alpha\dot\beta} \approx0$ this
means that our field $h^{\alpha(2)\dot\alpha(2)}$ describes exactly
five physical degrees of freedom.

Thus we have
\begin{equation}
W^{\alpha(4)} = \frac{1}{4} D^\alpha{}_{\dot\alpha}
\omega^{\alpha(3)\dot\alpha}, \qquad
B^{\alpha(3)\dot\alpha} = M \omega^{\alpha(3)\dot\alpha}, \qquad
\pi^{\alpha(2)\dot\alpha} = M^2 h^{\alpha(2)\dot\alpha}
\end{equation}
which describe first and second derivatives of our physical field that
do not vanish on-shell. However, to investigate all possible
interactions we need even higher derivatives. A straightforward way
to get all higher derivatives of a physical field which do not vanish
on-shell is by solving the unfolded equations. To see how this is
related to the results above, let us  calculate the next derivatives
of our three objects:
\begin{eqnarray}
D W^{\alpha(4)} &=& e_{\beta\dot\alpha} W^{\alpha(4)\beta\dot\alpha}
- \frac{M}{5} e^\alpha{}_{\dot\alpha} B^{\alpha(3)\dot\alpha},
\nonumber \\
D B^{\alpha(3)\dot\alpha} &=& e_{\beta\dot\beta}
B^{\alpha(3)\beta\dot\alpha\dot\beta} - \frac{M}{2} 
e_\beta{}^{\dot\alpha} W^{\alpha(3)\beta} - \frac{M}{4}
e^\alpha{}_{\dot\beta}\pi^{\alpha(2)\dot\alpha\dot\beta}, \\
D \pi^{\alpha(2)\dot\alpha(2)} &=& e_{\beta\dot\beta}
\pi^{\alpha(2)\beta\dot\alpha(2)\dot\beta} - \frac{M}{3}
e_\beta{}^{\dot\alpha} B^{\alpha(2)\beta\dot\alpha} - \frac{M}{3}
e^\alpha{}_{\dot\beta} B^{\alpha\dot\alpha(2)\dot\beta}, \nonumber
\end{eqnarray}
where
\begin{equation}
W^{\alpha(5)\dot\alpha} = \frac{1}{5} D^{\alpha\dot\alpha}
W^{\alpha(4)}, \qquad
B^{\alpha(4)\dot\alpha(2)} = \frac{1}{8} D^{\alpha\dot\alpha}
B^{\alpha(3)\dot\alpha}, \qquad
\pi^{\alpha(3)\dot\alpha(3)} = \frac{1}{9} D^{\alpha\dot\alpha}
\pi^{\alpha(2)\dot\alpha(2)}.  
\end{equation}
As a natural generalization of these results let us introduce the
following ansatz:
\begin{eqnarray}
W^{\alpha(4+k)\dot\alpha(k)} &=& \frac{4!}{(k+4)!k!}
(D^{\alpha\dot\alpha})^k W^{\alpha(4)}, \nonumber \\
B^{\alpha(3+k)\dot\alpha(k+1)} &=& \frac{3!}{(k+3)!(k+1)!}
(D^{\alpha\dot\alpha})^k B^{\alpha(3)\dot\alpha}, \label{sol_2m} \\
\pi^{\alpha(2+k)\dot\alpha(k+2)} &=& \frac{4}{(k+2)!^2}
(D^{\alpha\dot\alpha})^k \pi^{\alpha(2)\dot\alpha(2)}. \nonumber
\end{eqnarray}
To find the unfolded equations they satisfy, we need their
properties (see Appendix):
\begin{eqnarray}
D^\alpha{}_{\dot\beta} W^{\alpha(4+k)\dot\alpha(k-1)\dot\beta} &=& 0,
\nonumber \\
D^\alpha{}_{\dot\beta} B^{\alpha(3+k)\dot\alpha(k)\dot\beta} &=& 
\frac{(k+4)}{(k+1)} M W^{\alpha(4+k)\dot\alpha(k)}, \\
D^\alpha{}_{\dot\beta} \pi^{\alpha(2+k)\dot\alpha(k+1)\dot\beta} &=&
\frac{2(k+3)}{(k+2)} M B^{\alpha(3+k)\dot\alpha(k+1)}, \nonumber
\end{eqnarray}
\begin{eqnarray}
D_\beta{}^{\dot\alpha} W^{\alpha(3+k)\beta\dot\alpha(k)} &=& 
\frac{4(k+1)}{(k+4)} M B^{\alpha(3+k)\dot\alpha(k+1)}, \nonumber \\
D_\beta{}^{\dot\alpha} B^{\alpha(2+k)\beta\dot\alpha(k+1)} &=& 
\frac{3(k+2)}{(k+3)} M \pi^{\alpha(2+k)\dot\alpha(k+2)}, \\
D_\beta{}^{\dot\alpha} \pi^{\alpha(1+k)\beta\dot\alpha(k+2)} &=& 
\frac{2(k+3)}{(k+2)} M B^{\alpha(1+k)\dot\alpha(k+3)}, \nonumber
\end{eqnarray}
\begin{eqnarray}
D_{\beta\dot\beta} W^{\alpha(3+k)\beta\dot\alpha(k-1)\dot\beta} &=&
- \frac{(k+5)}{(k+4)} M^2 W^{\alpha(3+k)\dot\alpha(k-1)} \nonumber \\
D_{\beta\dot\beta} B^{\alpha(2+k)\beta\dot\alpha(k)\dot\beta} &=&
- \frac{k(k+5)}{(k+1)(k+3)} M^2 B^{\alpha(2+k)\dot\alpha(k)} \\
D_{\beta\dot\beta} \pi^{\alpha(1+k)\beta\dot\alpha(k+1)\dot\beta} &=&
- \frac{k(k+5)}{(k+2)^2} M^2 \pi^{\alpha(1+k)\dot\alpha(k+1)}.
\nonumber
\end{eqnarray}
Using these properties and the general structure (\ref{eq_gen}) one
can check that the resulting equations coincide with those previously
given  in (\ref{eq_2m}), so the expressions (\ref{sol_2m}) are
indeed the solution to these equations.

\section{Massive spin 5/2}

A massive spin 5/2 in four dimensions has six helicities 
$(\pm 5/2, \pm 3/2, \pm 1/2$ so a frame-like gauge invariant
description requires three physical fields\footnote{Here, as in all
fermionic cases, it is not necessary to introduce any auxiliary fields
to construct the Lagrangian.}: one-forms 
$\Phi^{\alpha(2)\dot\alpha}+h.c.$, $\Phi^\alpha+h.c.$ and zero-form 
$\phi^\alpha+h.c.$. The Lagrangian looks like:
\begin{eqnarray} 
{\cal L}_0 &=& - D \Phi_{\alpha\beta\dot\alpha} e^\beta{}_{\dot\beta}
\Phi^{\alpha\dot\alpha\dot\beta} + D \Phi_\alpha 
e^\alpha{}_{\dot\alpha} \Phi^{\dot\alpha}- 16M^2 D \phi_\alpha
E^\alpha{}_{\dot\alpha} \phi^{\dot\alpha} \nonumber \\
 && + \frac{M}{2} [ 3 \Phi_{\alpha\beta\dot\alpha} E^\beta{}_\gamma
\Phi^{\alpha\gamma\dot\alpha} - \Phi_{\alpha(2)\dot\alpha}
E^{\dot\alpha}{}_{\dot\beta} \Phi^{\alpha(2)\dot\beta}] 
 + 2a_0 \Phi_{\alpha(2)\dot\alpha} E^{\alpha(2)}
\Phi^{\dot\alpha} \nonumber \\
 && - 3M \Phi_\alpha E^\alpha{}_\beta \Phi^\beta + 32M^2
\Phi_\alpha E^\alpha{}_{\dot\alpha} \phi^{\dot\alpha}
+ 48M^3 E \phi_\alpha \phi^\alpha + h.c. \label{lag_m}
\end{eqnarray}
where
\begin{equation}
a_0{}^2 = \frac{5}{4}M^2.
\end{equation}
This Lagrangian is invariant under the following local gauge
transformations
\begin{eqnarray}
\delta \Phi^{\alpha(2)\dot\alpha} &=& D \rho^{\alpha(2)\dot\alpha}
+ e_\beta{}^{\dot\alpha} \rho^{\alpha(2)\beta} + \frac{M}{2}
e^\alpha{}_{\dot\beta} \rho^{\alpha\dot\alpha\dot\beta} + 
\frac{a_0}{3} e^{\alpha\dot\alpha} \rho^\alpha, \nonumber \\
\delta \Phi^\alpha &=& D \rho^\alpha + a_0 e_{\beta\dot\alpha}
\rho^{\alpha\beta\dot\alpha} + \frac{3M}{2} e^\alpha{}_{\dot\alpha}
\rho^{\dot\alpha}, \label{gauge_1} \\
\delta \phi^\alpha &=& \rho^\alpha. \nonumber
\end{eqnarray}
However, to construct a complete set of gauge invariant objects
(curvatures) we need three extra fields: one-form 
$\Phi^{\alpha(3)} + h.c.$ and zero-forms $\phi^{\alpha(3)} + h.c.$, 
$\phi^{\alpha(2)\dot\alpha} + h.c.$ with gauge transformations
\begin{equation}
\delta \Phi^{\alpha(3)} = D \rho^{\alpha(3)} + \frac{M^2}{3} 
e^\alpha{}_{\dot\alpha} \rho^{\alpha(2)\dot\alpha}, \qquad
\delta \phi^{\alpha(3)} = \rho^{\alpha(3)}, \qquad
\delta \phi^{\alpha(2)\dot\alpha} = \rho^{\alpha(2)\dot\alpha}.
\label{gauge_2}
\end{equation}
Note that here we also have a one-to-one correspondence between gauge
fields (one-forms) and Stueckelberg fields (zero-forms). Moreover,
each field has its own gauge invariant object (curvature). For the
gauge fields they are two-forms:
\begin{eqnarray}
{\cal F}^{\alpha(3)} &=& D \Phi^{\alpha(3)} + \frac{M^2}{3} 
e^\alpha{}_{\dot\alpha} \Phi^{\alpha(2)\dot\alpha} 
- \frac{2M^2}{3} E^\alpha{}_\beta \phi^{\alpha(2)\beta} 
- \frac{4M^2a_0}{9} E^{\alpha(2)} \phi^\alpha, \nonumber \\
{\cal F}^{\alpha(2)\dot\alpha} &=& D \Phi^{\alpha(2)\dot\alpha} + 
e_\beta{}^{\dot\alpha} \Phi^{\alpha(2)\beta} + \frac{M}{2}
e^\alpha{}_{\dot\beta} \Phi^{\alpha\dot\alpha\dot\beta} 
+ \frac{a_0}{3} e^{\alpha\dot\alpha} \Phi^\alpha, \\
{\cal F}^\alpha &=& D \Phi^\alpha + a_0 e_{\beta\dot\alpha}
\Phi^{\alpha\beta\dot\alpha} + \frac{3M}{2} 
e^\alpha{}_{\dot\alpha} \Phi^{\dot\alpha} 
- \frac{16M^2}{3} E^\alpha{}_\beta \phi^\beta
- 2a_0 E_{\beta(2)} \phi^{\alpha\beta(2)}, \nonumber
\end{eqnarray}
while for the Stueckelberg fields they are one-forms:
\begin{eqnarray}
{\cal C}^{\alpha(3)} &=& D \phi^{\alpha(3)} - \Phi^{\alpha(3)}
+ \frac{M^2}{3} e^\alpha{}_{\dot\alpha} 
\phi^{\alpha(2)\dot\alpha}, \nonumber \\
{\cal C}^{\alpha(2)\dot\alpha} &=& D \phi^{\alpha(2)\dot\alpha} -
\Phi^{\alpha(2)\dot\alpha} + \frac{M}{2}  
e^\alpha{}_{\dot\beta} \phi^{\alpha\dot\alpha\dot\beta} + 
e_\beta{}^{\dot\alpha} \phi^{\alpha(2)\beta} + \frac{a_0}{3}
e^{\alpha\dot\alpha} \phi^\alpha, \\
{\cal C}^\alpha &=& D \phi^\alpha - \Phi^\alpha + \frac{3M}{2}
e^\alpha{}_{\dot\alpha} \phi^{\dot\alpha} + a_0 
e_{\beta\dot\alpha} \phi^{\alpha\beta\dot\alpha}. \nonumber
\end{eqnarray}

The on-shell constraints look like:
\begin{equation}
{\cal F}^{\alpha(2)\dot\alpha} \approx 0, \qquad
{\cal F}^\alpha \approx 0, \qquad
{\cal C}^\alpha \approx 0.
\end{equation}
\begin{eqnarray}
0 &\approx& {\cal F}^{\alpha(3)} - E_{\beta(2)}
Y^{\alpha(3)\beta(2)},\nonumber \\
0 &\approx& {\cal C}^{\alpha(3)} - e_{\beta\dot\alpha}
\phi^{\alpha(3)\beta\dot\alpha}, \\
0 &\approx& {\cal C}^{\alpha(2)\dot\alpha} - e_{\beta\dot\beta}
\tilde\phi^{\alpha(2)\beta\dot\alpha\dot\beta}, \nonumber
\end{eqnarray}
where $Y^{\alpha(5)}$, $\phi^{\alpha(4)\dot\alpha}$ and
$\tilde\phi^{\alpha(3)\dot\alpha(2)}$ are gauge invariant zero-forms
which are just the first representatives of the infinite chains of
gauge invariant zero-forms satisfying the following unfolded
equations $(0 \le k < \infty)$:
\begin{eqnarray}
0 &=& D Y^{\alpha(5+k)\dot\alpha(k)} - e_{\beta\dot\beta}
Y^{\alpha(5+k)\beta\dot\alpha(k)\dot\beta} 
 + \frac{5M^2}{(k+5)(k+6)} e^\alpha{}_{\dot\beta} 
\phi^{\alpha(4+k)\dot\alpha(k)\dot\beta} \nonumber \\
 && + \frac{M^2}{(kk+1)(k+5)} e^{\alpha\dot\alpha} 
Y^{\alpha(4+k)\dot\alpha(k-1)}, \nonumber \\
0 &=& D \phi^{\alpha(4+k)\dot\alpha(k+1)} - e_{\beta\dot\beta}
\phi^{\alpha(4+k)\beta\dot\alpha(k+1)\dot\beta}
 + \frac{1}{(k+1)(k+2)} e_\beta{}^{\dot\alpha} 
Y^{\alpha(4+k)\beta\dot\alpha(k)} \label{eq_25m} \\
 && + \frac{4M^2}{(k+4)(k+5)} e^\alpha{}_{\dot\beta}
\tilde\phi^{\alpha(3+k)\dot\alpha(k+1)\dot\beta}
 + \frac{k(k+6)M^2}{(k+1)(k+2)(k+4)(k+5)}
e^{\alpha\dot\alpha} \phi^{\alpha(3+k)\dot\alpha(k)},  \nonumber \\
0 &=& D \tilde\phi^{\alpha(3+k)\dot\alpha(k+2)} - e_{\beta\dot\beta}
\tilde\phi^{\alpha(3+k)\beta\dot\alpha(k+2)\dot\beta}
 + \frac{2}{(k+2)(k+3)} e_\beta{}^{\dot\alpha} 
\phi^{\alpha(3+k)\beta\dot\alpha(k+1)} \nonumber \\
 && + \frac{3M}{(k+3)(k+4)} e^\alpha{}_{\dot\beta} 
\tilde{\phi}^{\alpha(2+k)\dot\alpha(k+2)\dot\beta} 
+ \frac{k(k+6)M^2}{(k+2)(k+3)^2(k+4)} e^{\alpha\dot\alpha}
\tilde\phi^{\alpha(2+k)\dot\alpha(k+1)}. \nonumber
\end{eqnarray}

In this case we also use an unitary gauge where all Stueckelberg
fields are set to zero:
\begin{equation}
\phi^{\alpha(3)} = 0, \qquad
\phi^{\alpha(2)\dot\alpha} = 0, \qquad
\phi^\alpha = 0.
\end{equation}
Then the on-shell constraints on the Stueckelberg curvatures give
\begin{equation}
\Phi^{\alpha(3)} \approx e_{\beta\dot\alpha} 
\phi^{\alpha(3)\beta\dot\alpha}, \qquad
\Phi^{\alpha(2)\dot\alpha} \approx e_{\beta\dot\beta}
\tilde\phi^{\alpha(2)\beta\dot\alpha\dot\beta}, \qquad
\Phi^\alpha \approx 0, \label{sol_2}
\end{equation}
while the gauge field curvatures take the form:
\begin{eqnarray}
{\cal F}^{\alpha(3)} &\approx& D \Phi^{\alpha(3)} + \frac{M^2}{3} 
e^\alpha{}_{\dot\alpha} \Phi^{\alpha(2)\dot\alpha}, \nonumber \\
{\cal F}^{\alpha(2)\dot\alpha} &\approx& D \Phi^{\alpha(2)\dot\alpha}
+ e_\beta{}^{\dot\alpha} \Phi^{\alpha(2)\beta} + \frac{M}{2}
e^\alpha{}_{\dot\beta} \Phi^{\alpha\dot\alpha\dot\beta},  \\
{\cal F}^\alpha &\approx& 0. \nonumber
\end{eqnarray}
Using (\ref{sol_2}) we obtain
\begin{eqnarray}
{\cal F}^{\alpha(2)\dot\alpha} &\approx& \frac{1}{2} E_{\beta(2)}
D^\beta{}_{\dot\beta} \tilde\phi^{\alpha(2)\beta\dot\alpha\dot\beta} -
E_{\beta(2)} \phi^{\alpha(2)\beta(2)\dot\alpha} \nonumber \\
 && + \frac{1}{2} E_{\dot\beta(2)} D_\beta{}^{\dot\beta} 
\tilde\psi^{\alpha(2)\beta\dot\alpha\dot\beta} - M E_{\dot\beta(2)}
\phi^{\alpha(2)\dot\alpha\dot\beta(2)}.
\end{eqnarray}
After (anti-)symmetrization on spinor indices this gives
\begin{eqnarray}
0 &\approx& \frac{1}{4} D^\alpha{}_{\dot\beta} 
\tilde\phi^{\alpha(3)\dot\alpha\dot\beta} - 
\phi^{\alpha(4)\dot\alpha}, \nonumber \\
0 &\approx& D_{\beta\dot\beta} 
\tilde\phi^{\alpha(2)\beta\dot\alpha\dot\beta}, \\
0 &\approx& \frac{1}{3} D_\beta{}^{\dot\alpha}
\tilde\phi^{\alpha(2)\beta\dot\alpha(2)} - M 
\phi^{\alpha(2)\dot\alpha(3)}.
\nonumber
\end{eqnarray}
Note that the last two equations mean that our physical field
$\tilde\phi^{\alpha(3)\dot\alpha(2)}+h.c.$ describes exactly six 
physical degrees of freedom. Similarly
\begin{equation}
{\cal F}^{\alpha(3)} \approx \frac{1}{2} E_{\beta(2)} 
D^\beta{}_{\dot\alpha} \phi^{\alpha(3)\beta\dot\alpha} + \frac{1}{2}
E_{\dot\alpha(2)} D_\beta{}^{\dot\alpha} 
\phi^{\alpha(3)\beta\dot\alpha} - M^2 E_{\dot\alpha(2)}
\tilde\phi^{\alpha(3)\dot\alpha(2)}.
\end{equation}
Thus we obtain
\begin{eqnarray}
0 &\approx& \frac{1}{5} D^\alpha{}_{\dot\alpha}
\phi^{\alpha(4)\dot\alpha} + Y^{\alpha(5)}, \nonumber \\
0 &\approx& D_{\beta\dot\alpha} \phi^{\alpha(3)\beta\dot\alpha}, \\
0 &\approx& \frac{1}{2} D_\beta{}^{\dot\alpha}
\phi^{\alpha(3)\beta\dot\alpha} - M^2 
\tilde\phi^{\alpha(3)\dot\alpha(2)}.
\nonumber
\end{eqnarray}
Note that from the last equation it follows that
$$
(\frac{1}{2} D^2 + M^2) \tilde\phi^{\alpha(3)\dot\alpha(2)} = 0.
$$
Thus we have 
\begin{equation}
Y^{\alpha(5)} = \frac{1}{5} D^\alpha{}_{\dot\alpha}
\phi^{\alpha(4)\dot\alpha}, \qquad
\phi^{\alpha(4)\dot\alpha} = \frac{1}{4} D^\alpha{}_{\dot\beta}
\tilde\phi^{\alpha(3)\dot\alpha\dot\beta}, \qquad
\tilde\phi^{\alpha(3)\dot\alpha(2)},
\end{equation}
which satisfy
\begin{eqnarray}
D_\beta{}^{\dot\alpha} Y^{\alpha(4)\beta} &=& M^2 
\phi^{\alpha(4)\dot\alpha}, \nonumber \\
D_\beta{}^{\dot\alpha} \phi^{\alpha(3)\beta\dot\alpha} &=& 
2M^2 \tilde\phi^{\alpha(3)\dot\alpha(2)}, \\
D_\beta{}^{\dot\alpha} \tilde\phi^{\alpha(2)\beta\dot\alpha(2)} &=&
3M \tilde\phi^{\alpha(2)\dot\alpha(3)}, \nonumber
\end{eqnarray}
and describe first and second derivatives of the physical field.

Now let us turn to the unfolded equations. Using the properties of our
three main objects we obtain
\begin{eqnarray}
D Y^{\alpha(5)} &=& e_{\beta\dot\beta} Y^{\alpha(5)\beta\dot\beta}
- \frac{M^2}{6} e^\alpha{}_{\dot\alpha} \phi^{\alpha(4)\dot\alpha}
\nonumber \\
D \phi^{\alpha(4)\dot\alpha} &=& e_{\beta\dot\beta}
\phi^{\alpha(4)\beta\dot\alpha\dot\beta} - \frac{M^2}{5}
e^\alpha{}_{\dot\beta} \tilde\phi^{\alpha(3)\dot\alpha\dot\beta}
- \frac{1}{2} e_\beta{}^{\dot\alpha} Y^{\alpha(4)\beta}, \\
D \tilde\phi^{\alpha(3)\dot\alpha(2)} &=& e_{\beta\dot\beta}
\tilde\phi^{\alpha(3)\beta\dot\alpha(2)\dot\beta} - \frac{M}{4}
e^\alpha{}_{\dot\beta} \tilde\phi^{\alpha(2)\dot\alpha(2)\dot\beta}
- \frac{1}{3} e_\beta{}^{\dot\alpha} \phi^{\alpha(3)\beta\dot\alpha},
\nonumber 
\end{eqnarray}
where
\begin{equation}
Y^{\alpha(6)\dot\alpha} = \frac{1}{6} D^{\alpha\dot\alpha}
Y^{\alpha(5)}, \qquad
\phi^{\alpha(5)\dot\alpha(2)} = \frac{1}{10} D^{\alpha\dot\alpha}
\phi^{\alpha(4)\dot\alpha}, \qquad
\tilde\phi^{\alpha(4)\dot\alpha(3)} = \frac{1}{12}
D^{\alpha\dot\alpha} \tilde\phi^{\alpha(3)\dot\alpha(2)}. 
\end{equation}
Now we introduce the following general ansatz:
\begin{eqnarray}
Y^{\alpha(5+k)\dot\alpha(k)} &=& \frac{5!}{(k+5)!k!}
(D^{\alpha\dot\alpha})^k Y^{\alpha(5)}, \nonumber \\
\phi^{\alpha(4+k)\dot\alpha(k+1)} &=& \frac{4!}{(k+4)!(k+1)!}
(D^{\alpha\dot\alpha})^k \phi^{\alpha(4)\dot\alpha}, \label{sol_25m} 
\\
\tilde\phi^{\alpha(3+k)\dot\alpha(k+2)} &=& \frac{3!2!}{(k+3)!(k+2)!}
(D^{\alpha\dot\alpha})^k \tilde\phi^{\alpha(3)\dot\alpha(2)}.
\nonumber
\end{eqnarray}
These new objects have the following properties:
\begin{eqnarray}
D^\alpha{}_{\dot\beta} Y^{\alpha(5+k)\dot\alpha(k-1)\dot\beta} &=& 0,
\nonumber \\
D^\alpha{}_{\dot\beta} \phi^{\alpha(4+k)\dot\alpha(k)\dot\beta} &=&
\frac{(k+5)}{(k+1)} Y^{\alpha(5+k)\dot\alpha(k)}, \\
D^\alpha{}_{\dot\beta} 
\tilde\phi^{\alpha(3+k)\dot\alpha(k+1)\dot\beta} &=& 
\frac{2(k+4)}{(k+2)} \phi^{\alpha(4+k)\dot\alpha(k+1)}, \nonumber 
\end{eqnarray}
\begin{eqnarray}
D_\beta{}^{\dot\alpha} Y^{\alpha(4+k)\beta\dot\alpha(k)} &=& 
\frac{5(k+1)}{(k+5)} M^2 \phi^{\alpha(4+k)\dot\alpha(k+1)}. \nonumber
\\
D_\beta{}^{\dot\alpha} \phi^{\alpha(3+k)\beta\dot\alpha(k+1)} &=&
\frac{4(k+2)}{(k+4)} M^2 \tilde\phi^{\alpha(3+k)\dot\alpha(k+2)}, \\
D_\beta{}^{\dot\alpha} \tilde\phi^{\alpha(2+k)\beta\dot\alpha(k+2)}
&=& 3M \tilde\phi^{\alpha(2+k)\dot\alpha(k+3)}, \nonumber
\end{eqnarray}
\begin{eqnarray}
D_{\beta\dot\beta} Y^{\alpha(4+k)\beta\dot\alpha(k-1)\dot\beta} &=&
- \frac{(k+6)}{(k+5)} M^2 Y^{\alpha(4+k)\dot\alpha(k-1)}, \nonumber \\
D_{\beta\dot\beta} \phi^{\alpha(3+k)\beta\dot\alpha(k)\dot\beta} &=& 
- \frac{k(k+6)}{(k+1)(k+4)} M^2  \phi^{\alpha(3+k)\dot\alpha(k)}, \\
D_{\beta\dot\beta} 
\tilde\phi^{\alpha(2+k)\beta\dot\alpha(k+1)\dot\beta} &=& 
- \frac{k(k+6)}{(k+2)(k+3)} M^2 
\tilde\phi^{\alpha(2+k)\dot\alpha(k+1)}. \nonumber
\end{eqnarray}
Using these properties and the formula (\ref{eq_gen}) one can check
that the resulting unfolded equations exactly coincide with
(\ref{eq_25m}) and so the expressions (\ref{sol_25m}) give their
solution.

\section{Integer spin}

To construct a gauge invariant Lagrangian for a massive boson with
spin $s$, we need a pair (physical field, auxiliary field) for each
helicity it contains: ($H^{\alpha(m)\dot\alpha(m)}$,
$\Omega^{\alpha(m+1)\dot\alpha(m-1)} + h.c.$), $1 \le m \le s-1$,
($A$, $B^{\alpha(2)} + h.c.$), ($\varphi$, $\pi^{\alpha\dot\alpha}$).
The general structure of the Lagrangian is typical of gauge
invariant descriptions of massive fields: a sum of the kinetic and
mass-like terms for all helicities plus cross terms
\begin{eqnarray}
\frac{1}{i} {\cal L}_0 &=& \sum_{m=1}^{s-1} (-1)^{m+1} 
[ (m+1) \Omega^{\alpha(m)\beta\dot\alpha(m-1)} E_\beta{}^\gamma
\Omega_{\alpha(m)\gamma\dot\alpha(m-1)} \nonumber \\
 && \qquad \qquad  - (m-1)
\Omega^{\alpha(m+1)\dot\alpha(m-2)\dot\beta} 
E_{\dot\beta}{}^{\dot\gamma} 
\Omega_{\alpha(m+1)\dot\alpha(m-2)\dot\gamma} \nonumber \\
 && \qquad \qquad +  2 \Omega^{\alpha(m)\beta\dot\alpha(m-1)}
e_\beta{}^{\dot\beta} D f_{\alpha(m)\dot\alpha(m-1)\dot\beta}] +
h.c. \nonumber \\
 && + [4E B_{\alpha(2)} B^{\alpha(2)} + 2 E_{\alpha(2)} B^{\alpha(2)}
D A + h.c.] - 6 E \pi_{\alpha\dot\alpha} \pi^{\alpha\dot\alpha} - 12
E_{\alpha\dot\alpha} \pi^{\alpha\dot\alpha} D \varphi \nonumber \\
&& + \sum_{m=2}^{s-1} (-1)^m a_m [ E_{\beta(2)}
\Omega^{\alpha(m-1)\beta(2)\dot\alpha(m-1)} 
f_{\alpha(m-1)\dot\alpha(m-1)} \nonumber \\
 && \qquad \qquad + \frac{(m-1)}{(m+1)} E_{\beta(2)}
f^{\alpha(m-2)\beta(2)\dot\alpha(m)}
\Omega_{\alpha(m-2)\dot\alpha(m)} + h.c. \nonumber \\
 && + a_1 [ \Omega^{\alpha(2)} E_{\alpha(2)} A - 2 B^{\beta\alpha}
E_\beta{}^{\dot\beta} f_{\alpha\beta} + h.c. ] + a_0
E_{\alpha\dot\alpha} \pi^{\alpha\dot\alpha} A \nonumber \\
 && + \sum_{m=1}^{s-1} (-1)^{m+1} b_m 
[ f^{\alpha(m-1)\beta\dot\alpha(m)} E_\beta{}^\gamma 
f_{\alpha(m-1)\gamma\dot\alpha(m)} + h.c. ] \nonumber \\
 && + \frac{a_1a_0}{2} E_{\alpha\dot\alpha}
f^{\alpha\dot\alpha} \varphi + 3a_1{}^2 E \varphi^2.
\end{eqnarray}
Here
\begin{eqnarray}
b_m &=& \frac{2s(s+1)}{m(m+1)(m+2)}M^2, \nonumber \\
a_m{}^2 &=& \frac{4(s-m)(s+m+1)}{m(m-1)} 
M^2 \qquad m \ge 2, \nonumber \\
a_1{}^2 &=& 2(s-1)(s+2) M^2, \\
a_0{}^2 &=& 24s(s+1)M^2. \nonumber 
\end{eqnarray}
This Lagrangian is invariant under the following gauge transformations
\begin{eqnarray}
\delta f^{\alpha(m)\dot\alpha(m)} &=&  D \xi^{\alpha(m)\dot\alpha(m)}
+ e_\beta{}^{\dot\alpha} \eta^{\alpha(m)\beta\dot\alpha(m-1)} +
e^\alpha{}_{\dot\beta} \eta^{\alpha(m-1)\dot\alpha(m)\dot\beta}
\nonumber \\
 && + \frac{m}{2(m+2)}a_{m+1} e_{\beta\dot\beta} 
\xi^{\alpha(m)\beta\dot\alpha(m)\dot\beta} + \frac{a_m}{2m(m+1)}
e^{\alpha\dot\alpha} \xi^{\alpha(m-1)\dot\alpha(m-1)}, \nonumber \\
\delta \Omega^{\alpha(m+1)\dot\alpha(m-1)} &=& D
\eta^{\alpha(m+1)\dot\alpha(m-1)} + \frac{a_{m+1}}{2} 
e_{\beta\dot\beta} \eta^{\alpha(m+1)\beta\dot\alpha(m-1)\dot\beta}
 + e_\beta{}^{\dot\alpha} \eta^{\alpha(m+1)\beta\dot\alpha(m-2)}
\nonumber \\
 &&  + \frac{b_m}{2(m+1)}  e^\alpha{}_{\dot\beta}
\xi^{\alpha(m)\dot\alpha(m-1)\dot\beta}
+ \frac{a_m}{2(m+1)(m+2)} e^{\alpha\dot\alpha}
\eta^{\alpha(m)\dot\alpha(m-2)},  \\
\delta f^{\alpha\dot\alpha} &=& D \xi^{\alpha\dot\alpha} 
+ e_\beta{}^{\dot\alpha} \eta^{\alpha\beta} + e^\alpha{}_{\dot\beta}
\eta^{\dot\alpha\dot\beta} + \frac{a_2}{6} e_{\beta\dot\beta}
\xi^{\alpha\beta\dot\alpha\dot\beta} - \frac{a_1}{4}
e^{\alpha\dot\alpha} \xi, \nonumber \\
\delta B^{\alpha(2)} &=& \frac{a_1}{2} \eta^{\alpha(2)}, \qquad
\delta A = D \xi - \frac{a_1}{2} e_{\alpha\dot\alpha}
\xi^{\alpha\dot\alpha}, \nonumber \\
\delta \pi^{\alpha\dot\alpha} &=& - \frac{a_1a_0}{24}
\xi^{\alpha\dot\alpha}, \qquad \delta \varphi = \frac{a_0}{12}.
\nonumber 
\end{eqnarray}

However, to construct a complete set of gauge invariant objects, we
need a set of extra one-forms that is just the sum of all extra
one-forms necessary for describing of massless fields with
helicities $3 \le h \le s$. In total, we must have
$\Omega^{\alpha(m+n)\dot\alpha(m-n)}$, $1 \le m \le s-1$,
$0 \le n \le m-1$, where $n=0,1$ correspond to the physical and
auxiliary fields. We also need an analogous set of Stueckelberg
zero-forms, The gauge invariant curvatures for the one-forms look
like:
\begin{eqnarray}
{\cal R}^{\alpha(m+n)\dot\alpha(m-n)} &=& D 
\Omega^{\alpha(m+n)\dot\alpha(m-n)} + e_\beta{}^{\dot\alpha}
\Omega^{\alpha(m+n)\beta\dot\alpha(m-n-1)} + a_{m,n}
e_{\beta\dot\beta} \Omega^{\alpha(m+n)\beta\dot\alpha(m-n)\dot\beta}
\nonumber \\
 && + b_{m,n} e^\alpha{}_{\dot\beta} 
\Omega^{\alpha(m+n-1)\dot\alpha(m-n)\dot\beta} + c_{m,n}
e^{\alpha\dot\alpha} \Omega^{\alpha(m+n-1)\dot\alpha(m-n-1)},
\end{eqnarray}
while for the Stueckelberg zero-forms we have
\begin{eqnarray}
{\cal C}^{\alpha(m+n)\dot\alpha(m-n)} &=& D 
\phi^{\alpha(m+n)\dot\alpha(m-n)} - 
\Omega^{\alpha(m+n)\dot\alpha(m-n)} + e_\beta{}^{\dot\alpha}
\phi^{\alpha(m+n)\beta\dot\alpha(m-n-1)} \nonumber \\
 && + a_{m,n} e_{\beta\dot\beta} 
\phi^{\alpha(m+n)\beta\dot\alpha(m-n)\dot\beta} + b_{m,n} 
e^\alpha{}_{\dot\beta} \phi^{\alpha(m+n-1)\dot\alpha(m-n)\dot\beta}
\nonumber \\
 && + c_{m,n} e^{\alpha\dot\alpha} 
\phi^{\alpha(m+n-1)\dot\alpha(m-n-1)}.
\end{eqnarray}
Here
\begin{eqnarray}
b_{m,n} &=& \frac{(s+n)(s-n+1)}{(m+n)(m+n+1)(m-n+1)(m-n+2)}
M^2, \nonumber \\
a_{m,n} &=& \frac{m(m+1)}{2(m-n+2)(m-n+1)}a_{m+1}, \\
c_{m,n} &=& \frac{1}{2(m+n)(m+n+1)}a_m. \nonumber
\end{eqnarray}

Now let us turn to the on-shell constraints. All gauge invariant
two-forms vanish except the highest one
\begin{equation}
{\cal R}^{\alpha(2s-2)} \approx E_{\beta(2)} 
\omega^{\alpha(s-2)\beta(2)}
\end{equation}
where the gauge invariant zero-form $\omega^{\alpha(2s)}$ is a
generalization of the Weyl tensor in gravity. Most of the gauge
invariant one-forms also vanish on-shell, the only non vanishing ones
being
\begin{equation}
{\cal C}^{\alpha(s-1+n)\dot\alpha(s-1-n)} \approx e_{\beta\dot\beta}
\omega^{\alpha(s-1+n)\beta\dot\alpha(s-1-n)\dot\beta}, \qquad 
0 \le n \le s-1.
\end{equation}
So we have a set of gauge invariant zero-forms 
$\omega^{\alpha(s+n)\dot\alpha(s-n)}$, $0 \le n \le s$ and again they
are just the first representatives of infinite chains of gauge
invariant zero-forms satisfying the unfolded equations 
($0 \le k < \infty$):
\begin{eqnarray}
0 &=& D \omega^{\alpha(2s+k)\dot\alpha(k)} - e_{\beta\dot\beta}
\omega^{\alpha(2s+k)\beta\dot\alpha(k)\dot\beta} 
 + \frac{2s}{(2s+k)(2s+k+1)} M^2 e^\alpha{}_{\dot\beta}
\omega^{\alpha(2s+k-1)\dot\alpha(k)\dot\beta} \nonumber \\
 && + \frac{1}{(k+1)(2s+k)} M^2 e^{\alpha\dot\alpha}
\omega^{\alpha(2s+k-1)\dot\alpha(k-1)}, \nonumber \\
0 &=& D \omega^{\alpha(s+n+k)\dot\alpha(s-n+k)} - e_{\beta\dot\beta}
\omega^{\alpha(s+n+k)\beta\dot\alpha(s-n+k)\dot\beta} \nonumber \\
 && + \frac{(s-n)(s-n-1)}{(s-n+k)(s-n+k+1)} e_\beta{}^{\dot\alpha}
\omega^{\alpha(s+n+k)\beta\dot\alpha(s-n+k-1)} \label{unf_bos} \\
 && + \frac{(s+n)}{(s-n)(s+n+k)(s+n+k+1)} M^2 e^\alpha{}_{\dot\beta}
\omega^{\alpha(s+n+k-1)\dot\alpha(s-n+k)\dot\beta} \nonumber \\
 && + \frac{k(2s+k+1)}{(s+n+k)(s+n+k+1)(s-n+k)(s+n+k+1)} M^2
e^{\alpha\dot\alpha} \omega^{\alpha(s+n+k-1)\dot\alpha(s-n+k-1)},
\nonumber \\
0 &=& D h^{\alpha(s+k)\dot\alpha(s+k)} - e_{\beta\dot\beta}
h^{\alpha(s+k)\beta\dot\alpha(s+k))\dot\beta} 
 + \frac{s(s-1)}{(s+k)(s+k+1)} e_\beta{}^{\dot\alpha}
\omega^{\alpha(s+k)\beta\dot\alpha(s-k-1)} \nonumber \\
 && + \frac{s}{(s-1)(s+k)(s+k+1)} e^\alpha{}_{\dot\beta}
\omega^{\alpha(s+k-1)\dot\alpha(s+k)\dot\beta} 
  + \frac{k(2s+k+1)}{(s+k)^2(s+k+1)^2} M^2
h^{\alpha(s+k-1)\dot\alpha(s-k-1)}. \nonumber
\end{eqnarray}

To find a solution for the on-shell constraints we again use the
unitary gauge, were all Stueckelberg  fields are set to zero. The the
on-shell constraints for the gauge invariant one-forms set to zero all
the one-forms with $m < s-1$ so we have only
\begin{equation}
\Omega^{\alpha(s-1+n)\dot\alpha(s-1-n)} = e_{\beta\dot\beta}
\omega^{\alpha(s-1+n)\beta\dot\alpha(s-1-n)\dot\beta}, \qquad
0 \le n \le s-1 \label{sol_3}
\end{equation}
and the remaining non-trivial on-shell constraints
\begin{eqnarray}
{\cal R}^{\alpha(2s-2)} &\approx& D \Omega^{\alpha(2s-2)} + b_{s-1}
e^\alpha{}_{\dot\alpha} \Omega^{\alpha(2s-3)\dot\alpha} \approx 
E_{\beta(2)} \omega^{\alpha(2s-2)\beta(2)}, \nonumber \\
{\cal R}^{\alpha(s-1+n)\dot\alpha(s-1-n)} &\approx& D
\Omega^{\alpha(s-1+n)\dot\alpha(s-1-n)} + e_\beta{}^{\dot\alpha}
\Omega^{\alpha(s-1+n)\beta\dot\alpha(s-2-n)} \nonumber \\
 && + b_{n}  e^\alpha{}_{\dot\beta} 
\Omega^{\alpha(s-2+n)\dot\alpha(s-1-n)\dot\beta} \approx 0, \\
{\cal T}^{\alpha(s-1)\dot\alpha(s-1)} &\approx&  D 
H^{\alpha(s-1)\dot\alpha(s-1)} + e_\beta{}^{\dot\alpha}
\Omega^{\alpha(s-1)\beta\dot\alpha(s-2)} + e^\alpha{}_{\dot\beta}
\Omega^{\alpha(s-2)\dot\alpha(s-1)\dot\beta} \approx 0, \nonumber
\end{eqnarray}
where
$$
b_n = \frac{M^2}{(s-1+n)(s-n)}.
$$
Using (\ref{sol_3}) we obtain
\begin{equation}
{\cal T}^{\alpha(s-1)\dot\alpha(s-1)} = \frac{1}{2} E_{\beta(2)}
D^\beta{}_{\dot\beta} h^{\alpha(s-1)\beta\dot\alpha(s-1)\dot\beta} -
(s-1) E_{\beta(2)} \omega^{\alpha(s-1)\beta(2)\dot\alpha(s-1)} + h.c.
\end{equation}
This gives
\begin{eqnarray}
0 &=& \frac{1}{(s+1)} D^\alpha{}_{\dot\beta}
h^{\alpha(s)\dot\alpha(s-1)\dot\beta} - (s-1)
\omega^{\alpha(s+1)\dot\alpha(s-1)}, \nonumber \\
0 &=& D_{\beta\dot\beta} h^{\alpha(s-1)\beta\dot\alpha(s-1)\dot\beta}.
\end{eqnarray}
Similarly
\begin{eqnarray}
{\cal R}^{\alpha(s-1+n)\dot\alpha(s-1-n)} &=& 
\frac{1}{2} E_{\beta(2)} D^\beta{}_{\dot\beta}
\omega^{\alpha(s-1+n)\beta\dot\alpha(s-1-n)\dot\beta} - (s-1-n)
E_{\beta(2)} \omega^{\alpha(s-1+n)\beta(2)\dot\alpha(s-1-n)} \nonumber
\\
 && + \frac{1}{2} E_{\dot\beta(2)} D_\beta{}^{\dot\beta}
\omega^{\alpha(s-1+n)\beta\dot\alpha(s-1-n)\dot\beta} 
- \frac{M^2}{(s-n)} E_{\dot\beta(2)}
\omega^{\alpha(s-1+n)\dot\alpha(s-1-n)\dot\beta(2)}
\end{eqnarray}
We obtain
\begin{eqnarray}
0 &=& \frac{1}{(s+1+n)} D^\alpha{}_{\dot\beta}
\omega^{\alpha(s+n)\dot\alpha(s-1-n)\dot\beta} - (s-1-n)
\omega^{\alpha(s+1+n)\dot\alpha(s-1-n)}, \nonumber \\
0 &=& D_{\beta\dot\beta} 
\omega^{\alpha(s-1+n)\beta\dot\alpha(s-1-n)\dot\beta}, \\
0 &=& \frac{1}{(s+1-n)} D_\beta{}^{\dot\alpha}
\omega^{\alpha(s-1+n)\beta\dot\alpha(s-n)} - \frac{M^2}{(s-n)}
\omega^{\alpha(s-1+n)\dot\alpha(s+1-n)}. \nonumber
\end{eqnarray}
From the last equation it follows that
$$
(\frac{1}{2} D^2 + M^2) \omega^{\alpha(s-1+n)\dot\alpha(s-1-n)}
\approx 0.
$$
At last
\begin{equation}
{\cal R}^{\alpha(2s-2)} =  \frac{1}{2} E_{\beta(2)} 
D^\beta{}_{\dot\alpha} \omega^{\alpha(2s-2)\beta\dot\alpha} 
+ \frac{1}{2} E_{\dot\alpha(2)} D_\beta{}^{\dot\alpha}
\omega^{\alpha(2-2)\beta\dot\alpha} - M^2 E_{\dot\alpha(2)}
\omega^{\alpha(2s-2)\dot\alpha(2)}.
\end{equation}
This gives
\begin{eqnarray}
0 &=& \frac{1}{2s} D^\alpha{}_{\dot\alpha}
\omega^{\alpha(2s-1)\dot\alpha} - \omega^{\alpha(2s)}, \nonumber \\
0 &=& D_{\beta\dot\alpha} \omega^{\alpha(2s-2)\beta\dot\alpha}, \\
0 &=& \frac{1}{2} D_\beta{}^{\dot\alpha}
\omega^{\alpha(2s-2)\beta\dot\alpha} - M^2
\omega^{\alpha(2s-2)\dot\alpha(2)}. \nonumber
\end{eqnarray}
Thus we have  objects describing up to $s$ derivatives of the physical
field $h^{\alpha(s)\dot\alpha(s)}$
\begin{eqnarray}
\omega^{\alpha(2s)} &=& \frac{1}{2s} D^\alpha{}_{\dot\alpha}
\omega^{\alpha(2s-1)\dot\alpha}, \nonumber \\
\omega^{\alpha(s+n)\dot\alpha(s-n)} &=&
\frac{1}{(s+n)(s-n)} D^\alpha{}_{\dot\beta}
\omega^{\alpha(s+n-1)\dot\alpha(s-n)\dot\beta}, \quad 1 \le n \le s-1
 \\
\omega^{\alpha(s)\dot\alpha(s)} &=& h^{\alpha(s)\dot\alpha(s)},
\nonumber
\end{eqnarray}
which satisfy
\begin{eqnarray}
D_\beta{}^{\dot\alpha} \omega^{\alpha(2s-1)\beta} &=& M^2
\omega^{\alpha(2s-1)\dot\alpha}, \nonumber \\
D_\beta{}^{\dot\alpha} \omega^{\alpha(s-1+n)\dot\alpha(s-1-n)} &=&
\frac{(s+1-n)}{(s-n)} M^2 
\omega^{\alpha(s-1+n)\dot\alpha(s+1-n)}, \\
D_\beta{}^{\dot\alpha} h^{\alpha(s-1)\beta\dot\alpha(s-1)} &=&
\frac{(s+1)}{(s-1)} \omega^{\alpha(s-1)\dot\alpha(s+1)}. \nonumber
\end{eqnarray}

We proceed with the solution of the unfolded equations. 
Using the properties given above we obtain
\begin{eqnarray}
D \omega^{\alpha(2s)} &=& e_{\beta\dot\alpha} 
\omega^{\alpha(2s)\beta\dot\alpha} - \frac{M^2}{(2s+1)} 
e^\alpha{}_{\dot\alpha} \omega^{\alpha(2s-1)\dot\alpha}, \nonumber \\
D \omega^{\alpha(s+n)\dot\alpha(s-n)} &=& e_{\beta\dot\beta} 
\omega^{\alpha(s+n)\beta\dot\alpha(s-n)\dot\beta} 
 - \frac{M^2}{(s+n+1)(s-n)} e^\alpha{}_{\dot\beta}  
\omega^{\alpha(s+n-1)\dot\alpha(s-n)\dot\beta} \nonumber \\
 && - \frac{(s-n-1)}{(s-n+1)} e_\beta{}^{\dot\alpha} 
\omega^{\alpha(s+n)\beta\dot\alpha(s-n-1)}, \\
D h^{\alpha(s)\dot\alpha(s)} &=& e_{\beta\dot\beta} 
h^{\alpha(s)\beta\dot\alpha(s)\dot\beta} 
- \frac{(s-1)}{(s+1)} e_\beta{}^{\dot\alpha} 
\omega^{\alpha(s)\beta\dot\alpha(s-1)} - \frac{(s-1)}{(s+1)}
e^\alpha{}_{\dot\beta} \omega^{\alpha(s-1)\dot\alpha(s)\dot\beta},
\nonumber
\end{eqnarray}
where
\begin{eqnarray}
\omega^{\alpha(2s+1)\dot\alpha} &=& \frac{1}{(2s+1)}
D^{\alpha\dot\alpha} \omega^{\alpha(2s)}, \nonumber \\
\omega^{\alpha(s+n+1)\dot\alpha(s-n+1)} &=& 
\frac{1}{(s+n+1)(s-n+1)} D^{\alpha\dot\alpha}
\omega^{\alpha(s+n)\dot\alpha(s-n)}, \\
h^{\alpha(s+1)\dot\alpha(s+1)} &=& \frac{1}{(s+1)^2}
D^{\alpha\dot\alpha} h^{\alpha(s)\dot\alpha(s)}. \nonumber
\end{eqnarray}
Now we introduce the following general ansatz
\begin{equation}
\omega^{\alpha(s+n+k)\dot\alpha(s-n+k)} = 
\frac{(s+n)!(s-n)!}{(s+n+k)!(s-n+k)!} (D^{\alpha\dot\alpha})^k
\omega^{\alpha(s+n)\dot\alpha(s-n)}, \qquad 0 \le n \le s.
\label{sol_b}
\end{equation}
These new objects have the following properties
\begin{eqnarray}
D^\alpha{}_{\dot\beta} \omega^{\alpha(2s+k)\dot\alpha(k-1)\dot\beta}
&=& 0 \nonumber \\
D^\alpha{}_{\dot\beta} 
\omega^{\alpha(s+n+k)\dot\alpha(s-n+k-1)\dot\beta} &=&
\frac{(s-n)(s-n-1)(s+n+k+1)}{(s-n+k)}
\omega^{\alpha(s+n+k+1)\dot\alpha(s-n+k-1)}, \\
D^\alpha{}_{\dot\beta} 
\omega^{\alpha(s+k)\dot\alpha(s+k-1)\dot\beta} &=&
\frac{s(s-1)(s+k+1)}{(s+k)}
\omega^{\alpha(s+k+1)\dot\alpha(s+k-1)}, \nonumber
\end{eqnarray}
\begin{eqnarray}
D_\beta{}^{\dot\alpha} \omega^{\alpha(2s+k-1)\beta\dot\alpha(k)} &=&
\frac{2s(k+1)}{(2s+k)} M^2 \omega^{\alpha(2s+k-1)\dot\alpha(k+1)},
\nonumber \\
D_\beta{}^{\dot\alpha} \omega^{\alpha(s+n+k-1)\beta\dot\alpha(s-n+k)}
&=& \frac{(s+n)(s-n+k+1)}{(s-n)(s+n+k)} M^2
\omega^{\alpha(s+n+k-1)\dot\alpha(s-n+k+1)}, \\
D_\beta{}^{\dot\alpha} h^{\alpha(s+k-1)\beta\dot\alpha(s+k)} &=&
\frac{s(s+k+1)}{(s-1(s+k)} \omega^{\alpha(s+k-1)\dot\alpha(s+k+1)},
\nonumber 
\end{eqnarray}
\begin{eqnarray}
D_{\beta\dot\beta} 
\omega^{\alpha(2s+k-1)\beta\dot\alpha(k-1)\dot\beta} &=& -
\frac{(2s+k+1)}{(2s+k)} M^2 \omega^{\alpha(2s+k-1)\dot\alpha(k-1)},
\nonumber \\
D_{\beta\dot\beta}
\omega^{\alpha(s+n+k-1)\beta\dot\alpha(s-n+k-1)\dot\beta} &=& -
\frac{k(2s+k+1)}{(s+n+k)(s-n+k)} M^2
\omega^{\alpha(s+n+k-1)\dot\alpha(s-n+k-1)}, \\
D_{\beta\dot\beta} h^{\alpha(s+k-1)\beta\dot\alpha(s+k-1)\dot\beta}
&=& - \frac{k(2s+k+1)}{(s+k)^2} M^2 
h^{\alpha(s+k-1)\dot\alpha(s+k-1)}.
\nonumber 
\end{eqnarray}
This in turn leads to the unfolded equations which coincide with
(\ref{unf_bos}) so that (\ref{sol_b}) is indeed their solution.

\section{Half-integer spin $s+\frac{1}{2}$}

In this case it is not necessary to introduce any auxiliary fields, so
to construct a Lagrangian we introduce a set of one-forms 
$\Phi^{\alpha(m+1)\dot\alpha(m)} + h.c.$, $0 \le m \le s-1$ and 
zero-form $\phi^\alpha + h.c.$. The free Lagrangian has the form
\begin{eqnarray}
{\cal L}_0 &=& \sum_{m=0}^{s-1} (-1)^{m+1} 
\Phi_{\alpha(m)\beta\dot\alpha(m)} e^\beta{}_{\dot\beta}
D \Phi^{\alpha(m)\dot\alpha(m)\dot\beta} - \phi_\alpha
E^\alpha{}_{\dot\alpha} D \phi^{\dot\alpha} \nonumber \\
 && + \sum_{m=1}^{s-1} (-1)^{m+1} c_m [ E^{\beta(2)}
\Phi_{\alpha(m-1)\beta(2)\dot\alpha(m)} 
\Phi^{\alpha(m-1)\dot\alpha(m)}] + c_0 \Phi_\alpha
E^\alpha{}_{\dot\alpha} \phi^{\dot\alpha} + h.c. \nonumber \\
 && + \sum_{m=0}^{s-1} (-)^{m+1} \frac{d_m}{2} [ (m+2) 
\Phi_{\alpha(m)\beta\dot\alpha(m)} E^\beta{}_\gamma
\Phi^{\alpha(m)\gamma\dot\alpha(m)} \nonumber \\
 && \qquad \qquad - m \Phi_{\alpha(m+1)\dot\alpha(m-1)\dot\beta}
E^{\dot\beta}{}_{\dot\gamma} 
\Phi^{\alpha(m+1)\dot\alpha(m-1)\dot\gamma}] + d_0 E
\phi_\alpha \phi^\alpha + h.c. 
\end{eqnarray}
Here
\begin{eqnarray}
d_m &=& \frac{(s+1)}{(m+1)(m+2)}M, \nonumber \\
c_m{}^2 &=& \frac{(s-m)(s+m+2)}{(m+1)^2}  M^2, \\
c_0{}^2 &=& 2s(s+2) M^2. \nonumber 
\end{eqnarray}
This Lagrangian is invariant under the following gauge transformations
\begin{eqnarray}
\delta \Phi^{\alpha(m+1)\dot\alpha(m)} &=& D 
\zeta^{\alpha(m+1)\dot\alpha(m)} + e_\beta{}^{\dot\alpha}
\zeta^{\alpha(m+1)\beta\dot\alpha(m-1)} + d_m
e^\alpha{}_{\dot\beta} \zeta^{\alpha(m)\dot\alpha(m)\dot\beta}
\nonumber \\
 && + c_{m+1} e_{\beta\dot\beta} 
\zeta^{\alpha(m+1)\beta\dot\alpha(m)\dot\beta} + 
\frac{c_m}{m(m+2)} e^{\alpha\dot\alpha}
\zeta^{\alpha(m)\dot\alpha(m-1)}, \\
\delta \phi^\alpha &=& c_0 \zeta^\alpha. \nonumber
\end{eqnarray}
Again, to construct a complete set of the gauge invariant objects we
have to introduce all extra one-forms corresponding to helicities $h
\ge 5/2$. So, in total we have $\Phi^{\alpha(m+n+1)\dot\alpha(m-n)} +
h.c.$, $1 \le m \le s-1$, $0 \le n \le m-1$, where $n=0$ corresponds
to physical fields. We also need a similar set of Stueckelberg 
zero-forms. The gauge invariant curvatures for the  one-forms look
like:
\begin{eqnarray}
{\cal F}^{\alpha(m+n+1)\dot\alpha(m-n)} &=& D 
\Phi^{\alpha(m+n+1)\dot\alpha(m-n)} + e_\beta{}^{\dot\alpha}
\Phi^{\alpha(m+n+1)\beta\dot\alpha(m-n-1)} + b_{m,n} 
e^\alpha{}_{\dot\beta} \Phi^{\alpha(m+n)\dot\alpha(m-n)\dot\beta}
\nonumber \\
 && + c_{m,n} e^{\alpha\dot\alpha} \Phi^{\alpha(m+n)\dot\alpha(m-n-1)}
+ a_{m,n} e_{\beta\dot\beta} 
\Phi^{\alpha(m+n+1)\beta\dot\alpha(m-n)\dot\beta},
\end{eqnarray}
while for the Stueckelberg zero-forms we have
\begin{eqnarray}
{\cal C}^{\alpha(m+n+1)\dot\alpha(m-n)} &=& D 
\phi^{\alpha(m+n+1)\dot\alpha(m-n)} - 
\Phi^{\alpha(m+n+1)\dot\alpha(m-n)} + e_\beta{}^{\dot\alpha}
\phi^{\alpha(m+n+1)\beta\dot\alpha(m-n-1)} \nonumber \\
 && + b_{m,n} e^\alpha{}_{\dot\beta} 
\phi^{\alpha(m+n)\dot\alpha(m-n)\dot\beta} + c_{m,n}
e^{\alpha\dot\alpha} \phi^{\alpha(m+n)\dot\alpha(m-n-1)} \nonumber \\
 && + a_{m,n} e_{\beta\dot\beta} 
\phi^{\alpha(m+n+1)\beta\dot\alpha(m-n)\dot\beta}.
\end{eqnarray}
Here
\begin{eqnarray}
a_{m,n} &=& \frac{(m+1)(m+2)}{(m-n+2)(m-n+1)}c_{m+1}, \nonumber \\
c_{m,n} &=& \frac{(m+1)}{m(m+n+1)(m+n+2)}c_m, \\
b_{m,n} &=& \frac{(s+n+1)(s-n+1)}{(m+n+1)(m+n+2)(m-n+1)(m-n+2)} M^2,  
\nonumber \\
b_{m,0} &=& d_m. \nonumber
\end{eqnarray}

Let us turn to the on-shell constraints. All gauge invariant
curvatures for the gauge fields vanish on-shell except the highest one
\begin{equation}
{\cal F}^{\alpha(2s-1)} \approx E_{\beta(2)} 
\psi^{\alpha(2s-1)\beta(2)}
\end{equation}
where $\psi^{\alpha(2s+1)}$ is a gauge invariant zero-form. At the
same time, most of the gauge invariant curvatures for the Stueckelberg
zero-forms also vanish, the only non-vanishing being
\begin{equation}
{\cal C}^{\alpha(s+n)\dot\alpha(s-n-1)} \approx e_{\beta\dot\beta} 
\psi^{\alpha(s+n)\beta\dot\alpha(s-n-1)\dot\beta}, \qquad
0 \le n \le s-1.
\end{equation}
Thus we have a set of gauge invariant zero-forms 
$\psi^{\alpha(s+n+1)\dot\alpha(s-n)}$, $0 \le n \le s$, which are just
the first representative of the infinite chains of the gauge invariant
zero-forms satisfying the following unfolded equations
\begin{eqnarray}
0 &=& D \psi^{\alpha(2s+k+1)\dot\alpha(k)} - e_{\beta\dot\beta}
\psi^{\alpha(2s+k)\beta\dot\alpha(k)\dot\beta} 
 + \frac{(2s+1)}{(2s+k+1)(2s+k+2)} M^2 e^\alpha{}_{\dot\beta}
\psi^{\alpha(2s+k)\dot\alpha(k)\dot\beta} \nonumber \\
 && + \frac{M^2}{(2s+k+1)(k+1)} e^{\alpha\dot\alpha}
 \psi^{\alpha(2s+k)\dot\alpha(k-1)}, \nonumber \\
0 &=& D \psi^{\alpha(s+n+k+1)\dot\alpha(s-n+k)} - e_{\beta\dot\beta}
\psi^{\alpha(s+n+k+1)\beta\dot\alpha(s-n+k)\dot\beta} \nonumber \\
 && + \frac{(s-n)(s-n-1)}{(s-n+k)(s-n+k+1)} e_\beta{}^{\dot\alpha} 
\psi^{\alpha(s+n+k+1)\beta\dot\alpha(s-n+k-1)} \nonumber \\
 && + \frac{(s+n+1)}{(s-n)(s+n+k+1)(s+n+k+2)} M^2 
e^\alpha{}_{\dot\beta} \psi^{\alpha(s+n+k)\dot\alpha(s-n+k)\dot\beta}
 \label{unf_fer} \\
 && + \frac{k(2s+k+2)}{(s+n+k+1)(s+n+k+2)(s-n+k)(s-n+k+1)} M^2
e^{\alpha\dot\alpha} \psi^{\alpha(s+n+k)\dot\alpha(s-n+k-1)},
\nonumber \\
0 &=& D \psi^{\alpha(s+k+1)\dot\alpha(s+k)} - e_{\beta\dot\beta}
\psi^{\alpha(s+k+1)\beta\dot\alpha(s+k)\dot\beta} 
 + \frac{s(s-1)}{(s+k)(s+k+1)} e_\beta{}^{\dot\alpha}
\psi^{\alpha(s+k+1)\beta\dot\alpha(s+k-1)} \nonumber \\
 && + \frac{(s+1)M}{(s+k+1)(s+k+2)} e^\alpha{}_{\dot\beta}
\psi^{\alpha(s+k)\dot\alpha(s+k)\dot\beta} \nonumber \\
 && + \frac{k(2s+k+2)}{(s+k)(s+k+1)^2(s+k+2)} M^2 e^{\alpha\dot\alpha}
\psi^{\alpha(s+k)\dot\alpha(s+k-1)}. \nonumber
\end{eqnarray}

We proceed with the solution of the on-shell constraints. After gauge
fixing (using again the unitary gauge) and solving relations on gauge
invariant one-forms we have only
\begin{equation}
\Phi^{\alpha(s+n)\dot\alpha(s-1-n)} = e_{\beta\dot\beta}
\psi^{\alpha(s+n)\beta\dot\alpha(s-1-n)\dot\beta}, \qquad
0 \le n \le s-1 \label{sol_4}
\end{equation}
and the remaining relations on the gauge invariant two-forms
\begin{eqnarray}
{\cal F}^{\alpha(2s-1)} &=& D \Phi^{\alpha(2s-1)} + \frac{M^2}{(2s-1)}
e^\alpha{}_{\dot\alpha} \Phi^{\alpha(2s-2)\dot\alpha}
 \approx E_{\beta(2)} \psi^{\alpha(2s-1)\beta(2)}, \nonumber \\
{\cal F}^{\alpha(s+n)\dot\alpha(s-1-n)} &=& D
\Phi^{\alpha(s+n)\dot\alpha(s-1-n)} + e_\beta{}^{\dot\alpha} 
\Phi^{\alpha(s+n)\beta\dot\alpha(s-2-n)} \nonumber \\
 && + \frac{M^2}{(s+n)(s-n)} e^\alpha{}_{\dot\beta}
\Phi^{\alpha(s-1+n)\dot\alpha(s-1-n)\dot\beta}  \approx 0, \\
{\cal F}^{\alpha(s)\dot\alpha(s-1)} &=& 
D \Phi^{\alpha(s)\dot\alpha(s-1)} + e_\beta{}^{\dot\alpha}
\Phi^{\alpha(s)\beta\dot\alpha(s-2)} + \frac{M}{s} 
e^\alpha{}_{\dot\beta} \Phi^{\alpha(s-1)\dot\alpha(s-1)\dot\beta}
\approx 0.  \nonumber
\end{eqnarray}
Using (\ref{sol_4}) we obtain
\begin{eqnarray}
{\cal F}^{\alpha(s)\dot\alpha(s-1)} &=& \frac{1}{2} E_{\beta(2)}
D^\beta{}_{\dot\beta} \psi^{\alpha(s)\beta\dot\alpha(s-1)\dot\beta} -
(s-1) E_{\beta(2)} \psi^{\alpha(s)\beta(2)\dot\alpha(s-1)} \nonumber
 \\
 && + \frac{1}{2} E_{\dot\beta(2)}D_\beta{}^{\dot\beta}
\psi^{\alpha(s)\beta\dot\alpha(s-1)\dot\beta} - M E_{\dot\beta(2)}
\psi^{\alpha(s)\dot\alpha(s-1)\dot\beta(2)}.
\end{eqnarray}
This gives
\begin{eqnarray}
0 &=& \frac{1}{(s+2)} D^\alpha{}_{\dot\beta}
\psi^{\alpha(s+1)\dot\alpha(s-1)\dot\beta} - (s-1)
\psi^{\alpha(s+2)\dot\alpha(s-1)}, \nonumber \\
0 &=& D_{\beta\dot\beta} 
\psi^{\alpha(s)\beta\dot\alpha(s-1)\dot\beta}, \\
0 &=& \frac{1}{(s+1)} D_\beta{}^{\dot\alpha}
\psi^{\alpha(s)\beta\dot\alpha(s)} - M
\psi^{\alpha(s)\dot\alpha(s+1)}. \nonumber
\end{eqnarray}
Similarly
\begin{eqnarray}
{\cal F}^{\alpha(s+n)\dot\alpha(s-n-1)} &=& \frac{1}{2} E_{\beta(2)}
D^\beta{}_{\dot\beta} 
\psi^{\alpha(s+n)\beta\dot\alpha(s-1-n)\dot\beta} - (s-1-n)
E_{\beta(2)} \psi^{\alpha(s+n)\beta(2)\dot\alpha(s-1-n)} \nonumber \\
 && + \frac{1}{2} E_{\dot\beta(2)} D_\beta{}^{\dot\beta}
\psi^{\alpha(s+n)\beta\dot\alpha(s-1-n)\dot\beta} 
- \frac{M^2}{(s-n)} E_{\dot\beta(2)}
\psi^{\alpha(s+n)\dot\alpha(s-1-n)\dot\beta(2)}.
\end{eqnarray}
We obtain
\begin{eqnarray}
0 &=& \frac{1}{(s+n+2)} D^\alpha{}_{\dot\beta}
\psi^{\alpha(s+n+1)\dot\alpha(s-n-1)\dot\beta} - (s-n-1)
\psi^{\alpha(s+n+2)\dot\alpha(s-n-1)}, \nonumber \\
0 &=& D_{\beta\dot\beta} 
\psi^{\alpha(s+n)\beta\dot\alpha(s-n-1)\dot\beta}, \\
0 &=& \frac{1}{(s-n+1)} D_\beta{}^{\dot\alpha}
\psi^{\alpha(s+n)\beta\dot\alpha(s-n)} - \frac{M^2}{(s-n)}
\psi^{\alpha(s+n)\dot\alpha(s-n+1)}. \nonumber
\end{eqnarray}
From the last equation it follows that
\begin{equation}
(\frac{1}{2} D^2 + M^2) \psi^{\alpha(s+n+1)\dot\alpha(s-n)} \approx 0.
\end{equation}
At last
\begin{equation}
{\cal F}^{\alpha(2s-1)} = \frac{1}{2} E_{\beta(2)} 
D^\beta{}_{\dot\alpha} \psi^{\alpha(2s-1)\beta\dot\alpha} + 
\frac{1}{2} E_{\dot\alpha(2)} D_\beta{}^{\dot\alpha}
\psi^{\alpha(2s-1)\beta\dot\alpha} - M^2 E_{\dot\alpha(2)}
\psi^{\alpha(2s-1)\dot\alpha(2)}.
\end{equation}
This gives
\begin{eqnarray}
0 &=& \frac{1}{(2s+1)} D^\alpha{}_{\dot\alpha}
\psi^{\alpha(2s)\dot\alpha} - \psi^{\alpha(2s+1)}, \nonumber \\
0 &=& D_{\beta\dot\alpha} \psi^{\alpha(2s-1)\beta\dot\alpha}, \\
0 &=& \frac{1}{2} D_\beta{}^{\dot\alpha}
\psi^{\alpha(2s-1)\beta\dot\alpha} - M^2
\psi^{\alpha(2s-1)\dot\alpha(2)}. \nonumber
\end{eqnarray}
Thus we have objects describing up to $s$ derivatives of the physical
field $\psi^{\alpha(s+1)\dot\alpha(s)}+h.c.$
\begin{eqnarray}
\psi^{\alpha(2s+1)} &=& \frac{1}{(2s+1)} D^\alpha{}_{\dot\alpha}
\psi^{\alpha(2s)\dot\alpha}, \\
\psi^{\alpha(s+n+1)\dot\alpha(s-n)} &=& \frac{1}{(s+n+1)(s-n)}
D^\alpha{}_{\dot\beta} \psi^{\alpha(s+n)\dot\alpha(s-n)\dot\beta},
\qquad 1 \le n \le s-1.  \nonumber 
\end{eqnarray}
They satisfy
\begin{eqnarray}
D_\beta{}^{\dot\alpha} \psi^{\alpha(2s)\beta} &=& M^2
\psi^{\alpha(2s)\dot\alpha}, \nonumber \\
D_\beta{}^{\dot\alpha} \psi^{\alpha(s+n)\beta\dot\alpha(s-n)} &=&
\frac{(s-n+1)}{(s-n)} M^2 \psi^{\alpha(s+n)\dot\alpha(s-n+1)}, \\
D_\beta{}^{\dot\alpha} \psi^{\alpha(s)\beta\dot\alpha(s)} &=&
(s+1) M \psi^{\alpha(s)\dot\alpha(s+1)}. \nonumber 
\end{eqnarray}
Now we turn to the unfolded equations. Using the properties given
above we obtain
\begin{eqnarray}
e_{\beta\dot\alpha} D^{\beta\dot\alpha} \psi^{\alpha(2s+1)} &=&
e_{\beta\dot\alpha} \psi^{\alpha(2s+1)\beta\dot\alpha} -
\frac{M^2}{(2s+2)} e^\alpha{}_{\dot\alpha} 
\psi^{\alpha(2s)\dot\alpha}, \nonumber \\
e_{\eta\dot\beta} D^{\beta\dot\beta} 
\psi^{\alpha(s+n+1)\dot\alpha(s-n)} &=& e_{\beta\dot\beta}
\psi^{\alpha(s+n+1)\beta\dot\alpha(s-n)\dot\beta} 
- \frac{M^2}{(s+n+2)(s-n)} e^\alpha{}_{\dot\beta}
\psi^{\alpha(s+n)\dot\alpha(s-n)\dot\beta} \nonumber \\
 && - \frac{(s-n-1)}{(s-n+1)}M^2 e_\beta{}^{\dot\alpha}
\psi^{\alpha(s+n+1)\dot\alpha(s-n-1)}, \\
e_{\beta\dot\beta} D^{\beta\dot\beta} \psi^{\alpha(s+1)\dot\alpha(s)}
&=& e_{\beta\dot\beta} \psi^{\alpha(s+1)\beta\dot\alpha(s)\dot\beta}
- \frac{M}{(s+2)} e^\alpha{}_{\dot\beta}
\psi^{\alpha(s)\dot\alpha(s)\dot\beta} - \frac{(s-1)}{(s+1)}
e_\beta{}^{\dot\alpha} \psi^{\alpha(s+1)\beta\dot\alpha(s-1)},
\nonumber 
\end{eqnarray}
where
\begin{eqnarray}
\psi^{\alpha(2s+2)\dot\alpha} &=& \frac{1}{(2s+2)}
D^{\alpha\dot\alpha} \psi^{\alpha(2s+1)}, \nonumber \\
\psi^{\alpha(s+n+2)\dot\alpha(s-n+1)} &=& 
\frac{1}{(s+n+2)(s-n+1)} D^{\alpha\dot\alpha} 
\psi^{\alpha(s+n+1)\dot\alpha(s-n)}, \\
\psi^{\alpha(s+2)\dot\alpha(s+1)} &=& \frac{1}{(s+1)(s+2)}
D^{\alpha\dot\alpha} \psi^{\alpha(s+1)\dot\alpha(s)}. \nonumber
\end{eqnarray}
Now we introduce the following general ansatz 
($0 \le n \le s$, $0 \le k < \infty$)
\begin{equation}
\psi^{\alpha(s+n+1+k)\dot\alpha(s-n+k)} = 
\frac{(s+n+1)!(s-n)!}{(s+n+1+k)!(s-n+k)!}
(D^{\alpha\dot\alpha})^k \psi^{\alpha(s+n+1)\dot\alpha(s-n)}.
 \label{sol_f}
\end{equation}
These new objects have the following properties
\begin{eqnarray}
D^\alpha{}_{\dot\beta} \psi^{\alpha(2s+1+k)\dot\alpha(k-1)\dot\beta}
&=& 0, \nonumber \\
D^\alpha{}_{\dot\beta} 
\psi^{\alpha(s+n+1+k)\dot\alpha(s-n+k-1)\dot\beta} &=& 
\frac{(s-n)(s-n-1)(s+n+k+2)}{(s-n+k)}
\psi^{\alpha(s+n+2+k)\dot\alpha(s-n-1+k)}, \nonumber \\
D^\alpha{}_{\dot\beta} \psi^{\alpha(s+1+k)\dot\alpha(s-1+k)\dot\beta}
&=& \frac{s(s-1)(2s+k+2)}{(s+k)} 
\psi^{\alpha(s+2+k)\dot\alpha(s-1+k)}, 
\end{eqnarray}
\begin{eqnarray}
D_\beta{}^{\dot\alpha} \psi^{\alpha(2s+k)\beta\dot\alpha(s+k)} &=&
\frac{(2s+1)(k+1)}{(2s+1+k)} M^2 \psi^{\alpha(2s)\dot\alpha},
\nonumber \\
 D_\beta{}^{\dot\alpha} 
\psi^{\alpha(s+n+k)\beta\dot\alpha(s-n+k)} &=& 
\frac{(s+n+1)(s-n+k+1)}{(s-n)(s+n+1+k)} M^2
\psi^{\alpha(s+n+k)\dot\alpha(s-n+k+1)}, \nonumber \\
D_\beta{}^{\dot\alpha} \psi^{\alpha(s+k)\beta\dot\alpha(s+k)} &=&
(s+1) M \psi^{\alpha(s+k)\dot\alpha(s+k+1)}, 
\end{eqnarray}
\begin{eqnarray}
D_{\beta\dot\beta} \psi^{\alpha(2s+k)\beta\dot\alpha(k-1)\dot\beta}
&=& - \frac{(2s+k+2)}{(2s+k+1)} M^2 
\psi^{\alpha(2s+k)\dot\alpha(k-1)}, \nonumber \\
D_{\beta\dot\beta} 
\psi^{\alpha(s+n+k)\beta\dot\alpha(s-n-1+k)\dot\beta} &&
- \frac{k(2s+k+2)}{(s+n+k+1)(s-n+k)} M^2 
\psi^{\alpha(s+n+k)\dot\alpha(s-n-1+k)}, \nonumber \\
D_{\beta\dot\beta} \psi^{\alpha(s+k)\beta\dot\alpha(s+k-1)\dot\beta}
&=& - \frac{k(2s+k+2)}{(s+k)(s+k+1)} M^2 
\psi^{\alpha(s+k)\dot\alpha(s+k-1)}
\end{eqnarray}
and this leads to the unfolded equations which coincide with
(\ref{unf_fer}) so that (\ref{sol_f}) is indeed their solution.

\section*{Conclusion}

In this paper, we filled some gap in the existing literature on higher
spins by presenting an explicit solution to the on-shell constraints
for a frame-like, gauge invariant description of massive, higher spin
fields in $d=4$. We began with the massive spin 2 and massive spin 5/2
as simple illustrations, and then considered arbitrary integer
and half-integer spin. We also showed that our results allow
us to find explicit solutions to the so-called unfolded equations
that determine all higher-order derivatives of the physical field
that are non-zero on-shell.

\appendix

\section{Appendix}

Restricting ourselves with $d=4$ case, where all objects correspond to
some (ir)reducible Lorentz group representations, let us consider the
following general ansatz for unfolded equations:
\begin{eqnarray}
0 &=& D \phi^{\alpha(k)\dot\alpha(l)} - e_{\beta\dot\beta}
\phi^{\alpha(k)\beta\dot\alpha(l)\dot\beta} + a_{k,l} 
e_\beta{}^{\dot\alpha} \phi^{\alpha(k)\beta\dot\alpha(l-1)} \nonumber
\\
 && + b_{k,l} e^\alpha{}_{\dot\beta}  
\phi^{\alpha(k-1)\dot\alpha(l)\dot\beta} + c_{k,l}
e^{\alpha\dot\alpha} \phi^{\alpha(k-1)\dot\alpha(l-1)}
\end{eqnarray}
Calculating one more derivative, taking into account that
$D \wedge  D = 0$ and expressing derivatives using again the same
equations, we obtain
\begin{eqnarray*}
0 &=&  [ - (l+2)c_{k+1,l+1} + lc_{k,l} - la_{k,l}b_{k+1,l-1} 
+ (l+2)a_{k-1,l+1}b_{k,l} ]
 E^\alpha{}_\beta \phi^{\alpha(k-1)\beta\dot\alpha(l)}  \\
 && + [ - (k+2)c_{k+1,l+1} + kc_{k,l} + (k+2)a_{k,l}b_{k+1,l-1} 
 - ka_{k-1,l+1}b_{k,l} ]
 E^{\dot\alpha}{}_{\dot\beta} \phi^{\alpha(k)\dot\alpha(l-1)\dot\beta}
  \\
 && + [la_{k,l} - (l+2)a_{k+1,l+1}] 
 E_{\beta(2)} \phi^{\alpha(k)\beta(2)\dot\alpha(l)} \\
 && + [- (k+2)b_{k+1,l+1} + kb_{k,l}]
 E_{\dot\beta(2)} \phi^{\alpha(k)\dot\alpha(l)\dot\beta(2)} \\
 && + [ 2(k+2)a_{k,l}c_{k+1,l-1} - 2ka_{k-1,l-1}c_{k,l}  ]
 E^{\dot\alpha(2)} \phi^{\alpha(k)\dot\alpha(l-2)} \\
 && + [ 2(l+2)c_{k-1,l+1}b_{k,l} - 2lc_{k,l}b_{k-1,l-1}  ]
 E^{\alpha(2)} \phi^{\alpha(k-2)\dot\alpha(l)} 
\end{eqnarray*}
So, for our ansatz to be consistent, all expressions within square
brackets must be zero. In fact, this system has an infinite number of
solutions, however, to find a meaningful explicit example, we need
some reasonable input, i.e. a set of fields to begin with. As a guide
on what kind of input is needed, let us once again use the 
decomposition into irreducible Lorentz group representations:
\begin{eqnarray}
e_{\beta\dot\beta} D^{\beta\dot\beta} \phi^{\alpha(k)\dot\alpha(l)} 
&=& \frac{1}{(k+1)(l+1)} [ e_{\beta\dot\beta} D^{(\beta|(\dot\beta}
\phi^{\alpha(k))|\dot\alpha(l))} - e^\alpha{}_{\dot\beta}
D_\beta{}^{(\dot\beta} \phi^{\alpha(k-1)\beta|\dot\alpha(l))}
\nonumber \\
 && \qquad \qquad - e_\beta{}^{\dot\alpha} D^{(\beta}{}_{\dot\beta}
\phi^{\alpha(k))\dot\alpha(l-1)\dot\beta} + e^{\alpha\dot\alpha}
D_{\beta\dot\beta} \phi^{\alpha(k-1)\beta\dot\alpha(l-1)\dot\beta}]
\label{eq_gen}
\end{eqnarray}
where round brackets denote symmetrization, Thus we must have
something like:
\begin{eqnarray}
\phi^{\alpha(k+1)\dot\alpha(l+1)} &\sim& D^{\alpha\dot\alpha}
\phi^{\alpha(k)\dot\alpha(l)}, \nonumber \\
\phi^{\alpha(k+1)\dot\alpha(l-1)} &\sim&  D^\alpha{}_{\dot\beta}
\phi^{\alpha(k)\dot\alpha(l-1)\dot\beta} \nonumber \\
\phi^{\alpha(k-1)\dot\alpha(l+1)} &\sim& D_\beta{}^{\dot\alpha}
\phi^{\alpha(k-1)\beta\dot\alpha(l)}, \\
\phi^{\alpha(k-1)\dot\alpha(l-1)} &\sim& D_{\beta\dot\beta}
\phi^{\alpha(k-1)\beta\dot\alpha(l-1)\dot\beta} \nonumber
\end{eqnarray}
In the main text we see that the on-shell conditions of the frame-like
formalism indeed provide us with the desired inputs.


\begin{thebibliography}{10}

\bibitem{Zin08b}
Yu.~M. Zinoviev
{\it "Frame-like gauge invariant formulation for massive high spin
particles",}
Nucl. Phys. {\bf B808} (2009) 185, arXiv:0808.1778.

\bibitem{PV10}
D.~S. Ponomarev, M.~A. Vasiliev
{\it "Frame-Like Action and Unfolded Formulation for Massive
Higher-Spin Fields",}
Nucl. Phys. {\bf B839} (2010) 466, arXiv:1001.0062.

\bibitem{KhZ19}
M.V. Khabarov, Yu.~M. Zinoviev
{\it "Massive higher spin fields in the frame-like multispinor
formalism",}
Nucl. Phys. {\bf B948} (2019) 114773, arXiv:1906.03438.

\bibitem{Zin24a}
Yu.~M. Zinoviev
{\it "On the Fradkin-Vasiliev formalism in d=4",}
Nucl. Phys. B {\bf 1012} (2025) 116839, arXiv:2410.16798.

\bibitem{Zin26}
Yu.~M. Zinoviev
{\it "On massive higher spins and gravity. IV. Arbitrary spin",}
  arXiv:2602.13661.

\bibitem{TV24}
Yu.~A. Tatarenko, M.~A. Vasiliev
{\it "Bilinear Fronsdal currents in the $AdS_{4}$ higher-spin
theory",}
JHEP {\bf 07} (2024) 246, arXiv:2405.02452.

\bibitem{BUV21}
A.S.Bychkov, K.A.Ushakov, M.A.Vasiliev
{\it "The $\sigma_-$ Cohomology Analysis for Symmetric Higher-Spin
Fields",}
Symmetry {\bf 13} (2021) 1498, arXiv:2107.01736.

\bibitem{BPSS15}
N.~Boulanger, D.~Ponomarev, E.~Sezgin, P.~Sundell
{\it "New Unfolded Higher Spin Systems in $AdS_3$",}
Class. Quant. Grav. {\bf 32} (2015) 155002, arXiv:1412.8209.

\bibitem{Zin15}
Yu.~M. Zinoviev
{\it "Massive higher spins in d=3 unfolded",}
J. Phys. A {\bf 49} (2016) 095401, arXiv:1509.00968.

\bibitem{BSZ16}
I.~L. Buchbinder, T.~V. Snegirev, Yu.~M. Zinoviev
{\it "Unfolded equations for massive higher spin supermultiplets in
$AdS_3$",}
JHEP {\bf 08} (2016) 075, arXiv:1606.02475.

\bibitem{KhZ20}
M.~V. Khabarov, Yu.~M. Zinoviev
{\it "Massive higher spin supermultiplets unfolded",}
Nucl. Phys. {\bf B953} (2020) 114959, arXiv:2001.07903.

\bibitem{Zin24}
Yu.~M. Zinoviev
{\it "On massive higher spin supermultiplets in d=4",}
JHEP {\bf 10} (2024) 222, arXiv:2408.08674.

\bibitem{KP18}
Sergei~M. Kuzenko, Michael Ponds
{\it "Topologically massive higher spin gauge theories",}
JHEP {\bf 10} (2018) 160, arXiv:1806.06643.

\bibitem{KhZ22}
M.~V. Khabarov, Yu.~M. Zinoviev
{\it "On massive higher spins in d=3",}
JHEP {\bf 04} (2022) 055, arXiv:2201.09491.

\end{thebibliography}
\end{document}